\documentstyle[proceedings,epsf]{crckapb}

\def\infig#1#2{\epsfysize=#1cm \centering{\mbox{\epsfbox{#2}}}}

\begin{opening}
\title{MAKING SPIRALS WITH COUNTER-ROTATING DISKS}
\subtitle{The case of NGC 4550}

\author{D. PFENNIGER}
\institute{Geneva Observatory, University of Geneva\\
           CH-1290 Sauverny, Switzerland}
\end{opening}

\runningtitle{MAKING SPIRALS WITH COUNTER-ROTATING DISKS}

\begin{document}

\begin{abstract}
A single merger scenario for making galaxies such as NGC 4550
possessing equal coplanar counter-rotating stellar disks is
investigated by collisionless N-body technique.  The scenario is
successful in producing an axisymmetric disk made of two almost equal
counter-rotating populations.  The final disk shows a clear bimodal
line profile in the outer part, which demonstrates that disk-disk
mergers do not always produce ellipticals.
\end{abstract}

\section{The Puzzle of NGC 4550}
The galaxy NGC 4550, an E7/S0 lenticular galaxy in the Virgo Cluster,
was discovered by Rubin et al. (1992) to be a galaxy consisting of two
coeval, coplanar, and counter-rotating stellar disks.  Many other
cases of counter-rotation are now known in ellipticals and also in
spirals, but what makes the case of NGC 4550 particular (Kuijken,
Fisher, \& Merrifield 1996) is that the mass ratio of the
counter-rotating disks is nearly $1/1$.  In the case of ellipticals
counter-rotation is so frequent (e.g.\ Schweizer 1998) that the merger
or accretion origins seem the most likely explanations for such
kinematic misalignments.

The difficulty with a merger scenario for NGC 4550 is that strong disk
mergers usually lead to ellipticals and the destruction of the disks,
a ``truth'' often believed to be general since the seminal paper of
Toomre \& Toomre (1972). Thakar \& Ryden (1996) have shown that over
several Gyr a series of well correlated small {\it gaseous\/} merger
events can lead to a massive counter-rotating gaseous disk.  But then
this scenario requires to preserve special correlations over several
Gyr, contrary to a single event.  Thus, it remains open whether a
single merger of two equal mass spirals can result in some
circumstances to a galaxy like NGC 4550.

In a forthcoming paper (Pfenniger \& Puerari 1998) we will present in
more details the general conditions leading to moderate heating.  In
addition to mass and energy, one must also consider the angular
momentum in the final budget. An input of specific angular momentum
tends to cool the system, while inputs of specific energy and mass
tend to heat it.

\section{Simulations}
\subsection{Merger scenario}
Since NGC 4550 is a rare object, a generic process is not required;
exceptional conditions can be acceptable and should be expected.
After various considerations we have retained: {\bf 1.} A nearly
circular orbit of the initial disks, supposing that the excess energy
of galaxies coming from infinity has already been absorbed by some
outer matter.  {\bf 2.}  Initial disks with opposite spins (see Fig.\
1).  Counter-rotating disks are ideal for minimizing shocks in outer
gaseous disks.  Also the retrograde disk is less affected by tidal
interactions.  {\bf 3.}  Nearly or exactly coplanar disks in the
orbital plane. This may look a rather improbable situation, but a
favorable factor to align the disk spins is to have initially
flattened dark matter (the torques on misaligned disks is then high).
Several arguments, such as the high frequency of warped outer HI disks
with straight line of nodes, do suggest flattened dark mass
distributions in spirals (see Pfenniger \& Combes 1994).

\subsection{Initial conditions and runs} 
The simulations presented here were run in Geneva. Independently,
Puerari run similar simulations in Mexico. We intend to publish
jointly both sets of simulations since they lead to the same
conclusions despite different choices of initial parameters.  In
particular Puerari's simulations include round massive halos, showing
that the shape of the halos does not change the qualitative results,
and his disks start on more elongated orbits.

In view of the poor understanding of gaseous processes in galaxies
leading to exaggerate viscosity with the SPH technique (see Thakar \&
Ryden 1996), we simulate first only the collisionless gravity part.
Obviously the resulting heating is enhanced with respect to 
simulations including gas.

Our two initial disks are identical, except for the initial
coordinates.  They consist of a bulge, with a scale-length of 1\,kpc,
an exponential disk with horizontal and vertical scale-lengths of 3
and 0.5\,kpc, and a flaring ($h_z = 0.03 R$) massive ``dark'' disk
with a constant surface density up to 10\,kpc, decreasing as $R^{-1}$
between 10 and 30\,kpc, and as $R^{-2}$ between 30 and 60\,kpc.  The
mass ratios of the three components are respectively (0.25:1:5), and
the particle masses are all equal.  The particle velocities are found by
solving the Jeans equation for each component in the total potential,
assuming velocity dispersions ellipsoids parallel to cylindrical
coordinates.  The potential of the mass distribution is calculated
with a set of $1.25\cdot 10^6$ particles on the Geneva PM polar grid
code.

A rapid run of coplanar counter-rotating disks merger is performed
with the PM code with $2.5\cdot 10^6$ particles.  The result is
positive but not fully convincing since during the merging process
substantial changes of positions of the mass distribution occur within
the grid, which has a position dependent resolution.  Therefore, we
select out a subset of $1.25\cdot10^5$ particles, the mass of which is
multiplied by a factor 20 to keep the same total mass.   The
particle subset is then run with the Barnes-Hut (1989) TREECODE (with an
opening angle of $\theta$=0.5), which does not imply any geometrical
assumptions.  In several experiments we verify that slight initial
disk inclinations with the orbital plane (5--$10^{\circ}$) still lead
to counter-rotating coplanar disks.

\begin{figure}
\vspace{-71truemm}
\hbox{\hspace{-9mm}\infig{14.1}{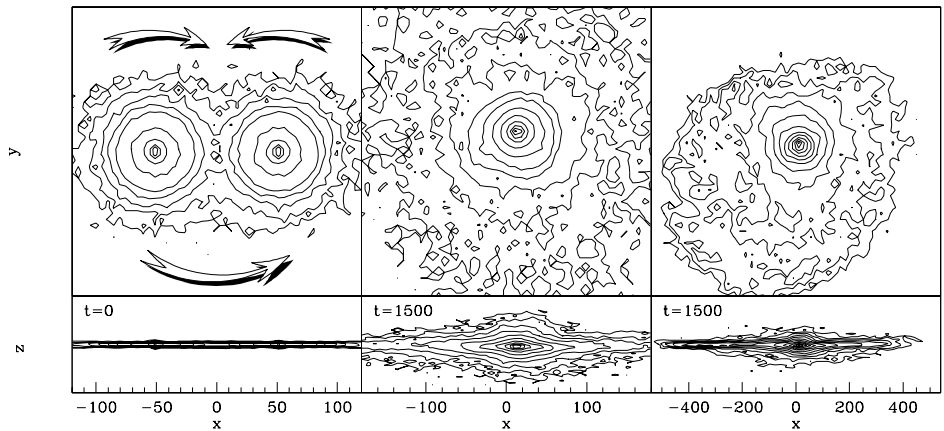}}
\vspace{-14truemm}
\caption{Merging of coplanar disks. The particle density is shown with
iso-contours separated by 2 mag. The arrows indicate the disk spins
and the disk sense of rotation.\hfill\break {\bf Left:} Initial conditions at
$t$=0\,Myr.\hfill\break 
{\bf Middle:} After the disk merging ($t$=1500\,Myr),
the remnant disk with counter-rotation.\hfill\break 
{\bf Right:} Large-scale view of the
particle distribution with marked asymmetries.}
\vspace{-3truemm}
\end{figure}

\subsection{Results}
Here we just describe the strictly coplanar disk merger.  Fig.\ 1,
left, shows the initial particle distribution with the senses of
rotation. The disks are on a prograde near circular orbit. The tidal
perturbation creates immediately (at $\sim$200\,Myr) a bar in each
disks, persisting until the disks merge.  The merging process
conserves the disks fairly well, the less damaged one being the
retrograde one.  Fig.\ 1, middle, shows the inner remnant disk at
$t$=1500\,Myr, still containing 80\% of the initial total mass, almost
circular and barless.  Fig.\ 1, right, shows the large-scale
particle distribution, at $t$=1500\,Myr.  The excess angular momentum
is transported by 10-20\% of the mass, mostly of the prograde
disk, to large distances, 100--500\,\rm kpc.

\begin{figure}  
\vspace{-3mm}  
\parbox[b]{9 cm}{\vspace{-6cm}\hspace{-3mm}\infig{9}{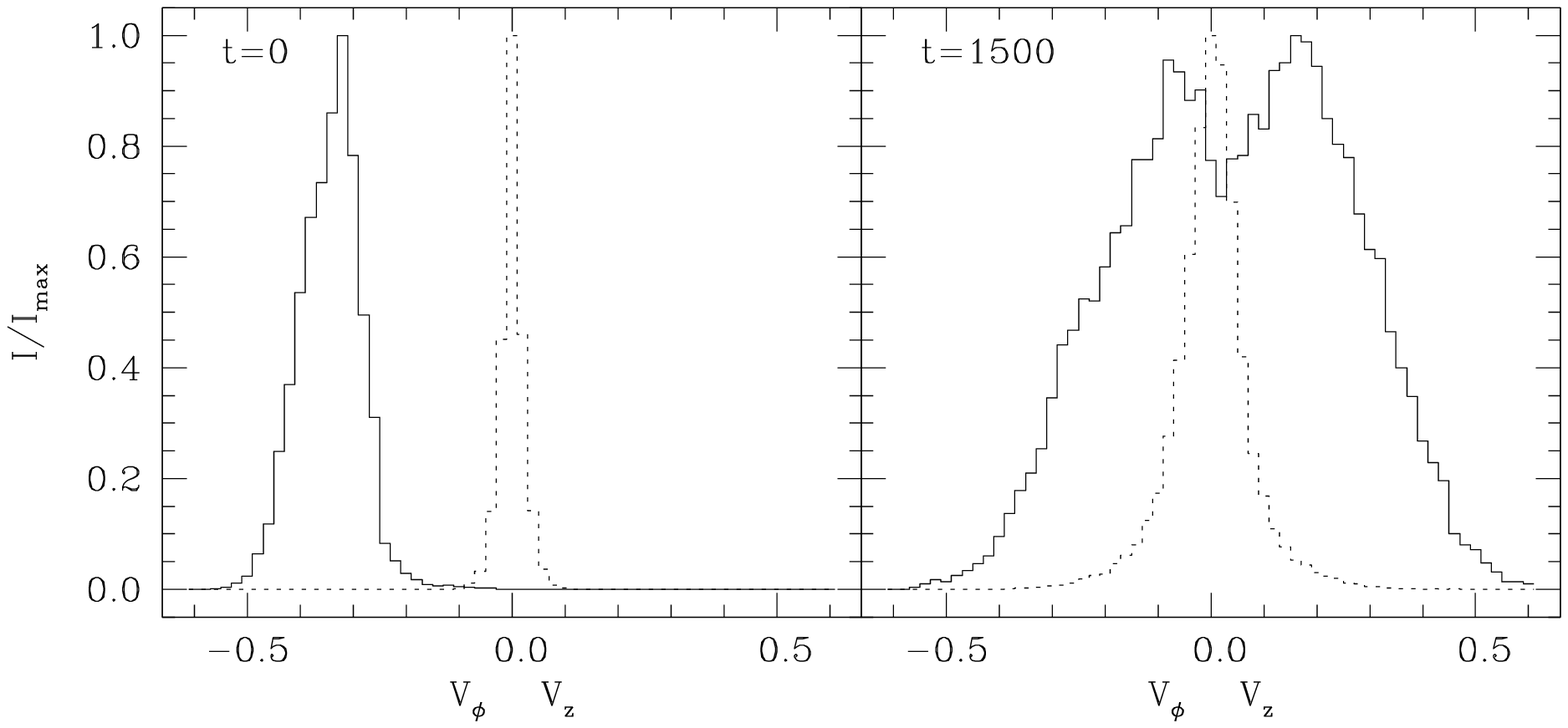}\vspace{-5mm}}\hfill\parbox[b]{3.5 cm}{
\caption{Tangential (solid) and vertical (dotted) line profiles,
averaged in the interval 30$<${}$R${}$<$100.\hfill\break
{\bf Left:} In the initial retrograde disk, 
{\bf Right:} In the final remnant. \hfill\break
The final bimodal tangential velocity dispersion is clearly visible.}}
\end{figure}

Fig.\ 2 shows how the average line profiles as seen in the edge-on
disks change from unimodal to bimodal.  The bimodal distribution is
crucial in order to distinguish counter-rotating disks from a hot
population with a zero net rotation. The ratios of final to initial 
velocity dispersion amount to $\sim$2.4 in the radial direction, and
$\sim$2 in the vertical direction.  The final mass distribution 
follows well a $R^{1/4}$ profile. With a relatively limited heating
the final disk resembles both in shape and kinematics a typical
S0/E7 galaxy, with, in addition, marked counter-rotating populations.

At some moments during the merging process the system as seen edge-on
looks like a single galaxy but with two bulges.  Such a case of
``two-bulge'' looking galaxy is PGC 57064 in the Hercules cluster.

\section{Conclusions}

We have shown that counter-rotating co-spatial disks can be made by a
single spiral-spiral merger.  The required initial conditions are
somewhat peculiar, but favored if the dark matter distribution is
flat; however, this is not a necessary condition.  The strong reaction
of the disks, expelling 10--20\% of the mass to 100--500\,kpc, means
that dark matter in such systems must be distributed in a pronounced
asymmetric way for several Gyr.  Thus, in many spirals the outer (dark)
mass distribution should still be chaotic.


\end{document}